\documentclass[oribibl,fullpage]{llncs}
\usepackage{multicol}
\usepackage{amsmath}
\usepackage{amssymb}
\usepackage{graphicx}
\usepackage{color}
\usepackage{tabu}

\usepackage{algorithm}
\usepackage{algpseudocode}
\usepackage{enumfrom}

\newcommand{\ignore}[1]{}
\newcommand{\boxtheorem}{\hfill $\blacksquare$\vspace{1mm}}

\newcounter{example-counter}
\renewenvironment{example}%
{\vskip \abovedisplayskip \refstepcounter{example-counter}%
\noindent {\bf Example \arabic{example-counter}.}}%

\newcounter{remark-counter}
\renewenvironment{remark}%
{\vskip \abovedisplayskip \refstepcounter{remark-counter}%
\noindent {\bf Remark \arabic{remark-counter}.}}%
{\boxtheorem}

\newcommand{\mbf}[1]{{\mathbf #1}}

\newcommand{\msf}[1]{\mbox{\sf #1}}

\newcommand{\red}[1]{\textcolor{red}{#1}}
\newcommand{\blue}[1]{\textcolor{blue}{#1}}

\newcommand{\nit}[1]{{\it #1}}
\newcommand{\mc}[1]{\mathcal{ #1}}

\newcommand{\defproof}[2]{{\noindent\bf Proof of #1:\
}#2 \boxtheorem \vspace{2mm}}
\makeatother

\newcommand{\msfbf}[1]{\mathsf{\bf #1}}

\newcommand{\set}[1]{\{#1\}}

\newcommand{\core}{\mathsf{Core}}%

\newcommand{\wrt}{\text{w.r.t.\ }}

 \newcommand{\comlb}[1]{{\vspace{2mm}\noindent \red{\bf COMM(LEO):}}~ #1 \hfill {\bf    END.}\\}
 
 \newcommand{\nina}[1]{{\vspace{2mm}\noindent \bf  {\blue{COMM(NINA):}}}~ #1 \hfill {\bf
     \blue{END.}}\vspace*{2mm} \\ }

\title{{\bf Sufficient Explanations in Databases and their Connections to Database Repairs}}

\author{{\bf Leopoldo Bertossi}\inst{1}\thanks{Emeritus Professor, Carleton University. } \and
{\bf Nina Pardal}\inst{2}}
\institute{Carleton University, Canada \& IMFD, Chile. \and University of Edinburgh, UK.}

\begin{document}

\maketitle
\pagestyle{plain}
\thispagestyle{empty}

\begin{abstract}
We investigate the notion of sufficient explanation, and a sufficiency-degree as attribution score for database tuples in relation to query answering.   We also investigate and exploit connections with database  repairs as used for dealing with inconsistent databases; and with causality-based necessary explanations, obtaining new computational results. We show how to use answer-set programs to specify sufficient explanations and compute sufficiency-degrees.
\end{abstract}

\section{Introduction} \label{sec:intro}

\vspace{-2mm}
In parallel to the developments of Explainable AI (XAI), there have been different proposals for Explainable Data Management (XDM), most commonly for providing explanations for query results. 

Early efforts were based on {\em data provenance}, which is still an active area of research \cite{BunemanKT01,tannen,ester24a,ester24b}.  
\ Independently,  the notion of causality-based explanation for a query result   was introduced in \cite{Meliou2010a}, on the basis of the deeper concepts of {\em counterfactual} and {\em actual cause} that can be traced back to
\cite{Halpern05}. Tuples of a database (DB) instance become explanations for query answers. Even more, 
they can be quantitatively ranked, using    {\em causal responsibility} \cite{Chockler04}, to quantify the extent by which a cause contributes to an answer \cite{tocs}. Responsibility becomes an {\em attribution score}, in the current terminology of XAI. 

\ignore{++
\vspace{-0.7cm}
\begin{figure}[h!]
\begin{center}
\includegraphics[width=2.2cm]{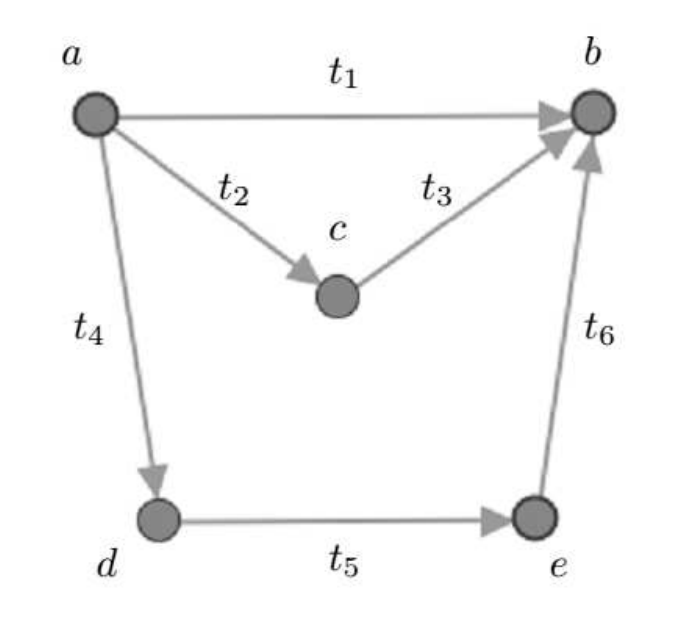}
\vspace{-3mm}\caption{A Graph DB.}\label{fig:first}
\end{center}
\end{figure}

++}

\vspace{-2mm}
\begin{example} \label{exa:newcrIntro} 
Instance $D$ is represented here as a directed graph, and also as a

\vspace{-4mm}
\begin{multicols}{2}
 \vspace*{-3mm}\hspace*{1.5cm}\includegraphics[width=2.2cm]{newcont.pdf}
 
\noindent  single relation $E = \{(t_1,a,b), (t_2,a,c),$ $ (t_3,c,b),$ $ 
(t_4,a,d), 
(t_5,d,e), $ $(t_6,e,b)\}$. \  $t_1$-$t_6$ label the edges in the graph, and become tuple
 identifiers (tids) in $D$.  
 \end{multicols}
 
 \vspace{-4mm}The Boolean query $\mc{Q}$ asks if there is a path connecting nodes $a$ and $b$.\footnote{It can be expressed  as the Datalog query:  $\nit{yes}\leftarrow P(a, b).$, $P(x, y) \leftarrow P(x, z) E(z, y).$, $P(x, y)\leftarrow E(x, y).$ }
The answer is positive, and all the tuples turn out to be actual causes for the positive answer, with causal responsibility $1/3$ (see Example \ref{ex:exIntro3} below).
\ Intuitively speaking for the moment, if we {\em counterfactually remove} any tuple $t$ from $E$, expecting to invalidate the positive query answer (QA), we have to further and counterfactually delete two extra tuples, that is, three necessary deletions in total, hence 1/3. Every tuple belongs to a three-element  {\em necessary set of tuples} for the positive answer, actually a {\em minimal} one under set inclusion: If we remove that set from $E$ (but not a proper subset), the answer is invalidated.
    \boxtheorem\end{example}

The example may look counterintuitive: $t_1$, the direct edge,
seems to contribute more than the other tuples, e.g. $t_2$ and $t_3$, which motivated  the use of the {\em causal-effect score} as an alternative attribution score \cite{tapp16}.

 {\em Sufficient explanations} have been applied in XAI, sometimes as {\em abductive explanations} \cite{marquis,darwiche,joao,izza}, but they have received much less attention from the DB community. \ In our case, sufficient explanations take the form of {\em sufficient-sets} of tuples, and tuples in them become individual explanations.
Intuitively, a tuple $t \in D$ is a sufficient explanation for $\mc{Q}$ being true in $D$ if $t$, possibly with some other tuples in $D$,  makes the query true.

\vspace{-1mm}
\begin{example} \label{ex:exIntro2}  (ex. \ref{exa:newcrIntro} cont.) Tuple $t_1$ suffices to make the query true: \ $\{t_1\} \models \mc{Q}$, so as the other ``small" subinstances  $\{t_2,t_3\}$ and $\{t_4,t_5, t_6\}$. \ All the tuples in $E$ are sufficient explanations, but some of them seem to be stronger than others:  $t_1$ alone suffices to make the query true. 
\boxtheorem\end{example}

\vspace{1mm}
In this work, we investigate {\em sufficient explanations} for QA in DBs, and introduce a numerical {\em sufficiency score} that is based on the latter. This is an attribution score, as a measure of the sufficiency of a tuple for a positive QA.
\ We also establish connections with {\em necessary explanations} that are defined as the above mentioned {\em actual causes}.    In this work we consider only Boolean queries. The extension to open queries is straightforward.

Given an inconsistent DB $D$ w.r.t. a set of integrity constraints (ICs), the {\em database repairs} are the database instances that are consistent and minimally depart from $D$ \cite{CQA}. 
\ A useful connection between  repairs and actual causes for QA was established  in \cite{tocs}, using the negation of the query as a {\em denial constraint}.  \ 

Without appealing to the relationship between necessary and sufficient explanations, we establish a direct connection between the latter and database repairs. 
 In particular, we show how the intersection of all repairs, which we call {\em the repair core}, or simply {\em the core}, can be used to efficiently compute sufficient explanations and sufficiency scores for DB tuples. The core can be computed and materialized once, independently from the explanations later considered; and then, reused across all explanation computations. 
 
More specifically, our contributions are as follows:

\vspace{1mm}
\noindent 1. \ We formalize the notion of tuple-based sufficient explanation for monotone Boolean queries in DBs. On this basis, we introduce the quantitative {\em sufficiency-score} for sufficient tuples. \ DBs  may contain  {\em endogenous and exogenous} tuples, as is common in causality scenarios; which also makes sense in data management. 


\noindent 2. \  We establish connections between necessary and sufficient explanations and the repair core, showing that tuples in the core are exactly those with necessity-degree $0$, and do not belong to any minimal sufficient-set. We use the core to provide a direct connection between sufficient explanations for conjunctive queries (CQs) and database repairs.


\noindent 3. \  Given a self-join-free query $\mc{Q}$, we exploit the efficient computation of the core to, for each tuple outside the core, efficiently compute a  minimum sufficient-set that contains it. With this, we can compute the  sufficiency degrees of those tuples.  All this in data complexity. 

\noindent 4. \ We show how to use {\em answer-set programming} (ASP) to specify sufficient explanations and compute sufficiency degrees. \ignore{We also represent this task as an instance of an {\em abductive problem}.}

\noindent 5. \  We discuss and elaborate on possibly fruitful directions of further research.

\vspace{1mm}
This paper is organized as follows.  In Section \ref{sec:pre}, we provide background material and a discussion of related work. In Section \ref{sec:sufficiency}, we introduce the
notion of  {\em sufficient  tuple} and the  {\em sufficient tuple} for QA and the quantitative {\em sufficiency degree} of a tuple.  In Section \ref{sec:repair-core}, we define the repair core and establish its connections to sufficient tuples. In Section \ref{sec:explCore}, we exploit that connection to compute sufficient explanations. In Section \ref{sec:asp}, we specify and compute sufficient explanations by means of ASPs. In Section \ref{sec:disc}, we end with a discussion and pointing to promising research directions.
Proofs are provided in the Appendix.

\vspace{-3mm}
\section{Preliminaries} \label{sec:pre}

\vspace{-2mm}
\paragraph{\bf Relational Databases.}
A relational database  (DB) is a finite set of ground atomic formulas, a.k.a. atoms or tuples, which will be denoted with $t, t_1, ...$.  \ A {\em conjunctive query}  (CQ) is a formula $\mc{Q}(\bar{x})$ of the form \ $\exists \bar{y}(P_1(\bar{s}_1) \wedge \cdots \wedge P_m(\bar{s}_m))$, with $\bar{x}$ showing
the free variables,   i.e. those not in $\bar{y}$.  If $\bar{x}$ is empty,   the query is {\em Boolean} (a BCQ),   i.e. the query is a sentence that is true or false
in a DB,   denoted by $D \models \mc{Q}$ and $D \not\models \mc{Q}$,   respectively.  \ If $\mc{Q}$ is a BCQ, the answer $\mc{Q}[D]$ belongs to $\{0,1\}$. The {\em size} of a CQ is its number of atoms.  A CQ is {\em self-join-free}, denoted sjf-\!CQ or sjf-\!BCQ, if it does not contain repeated predicate symbols (the $P_i$s). \ A query $\mc{Q}$ is {\em monotone} if for every two instances $D_1 \subseteq D_2$, \ $\mc{Q}[D_1] \subseteq \mc{Q}[D_2]$; and, for a BCQ, $\mc{Q}[D_1] \leq \mc{Q}[D_2]$. That is, the set of answers grows monotonically with the DB.   {CQ}s, unions of  {CQs}, and Datalog queries are monotone. We consider only monotone queries (MQs); if Boolean, we write BMQ.

\vspace{-2mm}
\paragraph{\bf Database  Repairs.}
For a BCQ
$\mc{Q}\!: \exists \bar{x}(P_1(\bar{x}_1) \wedge \cdots \wedge P_m(\bar{x}_m))$, \
$\neg \mc{Q}$ \ becomes a {\em denial constraint} (DC): \
$\mc{K}(\mc{Q})\!: \   \neg \exists \bar{x}(P_1(\bar{x}_1) \wedge \cdots \wedge P_m(\bar{x}_m))$. \ $\mc{Q}$ holds in $D$ \  iff \  $D$ is inconsistent w.r.t. $\mc{K}(\mc{Q})$, in which case we may consider {\em database repairs} \cite{CQA}. {\em Subset-repairs}, a.k.a. {\em S-repairs}, are subset-maximal consistent subinstances of $D$.
\ {\em Cardinality repairs}, a.k.a.  {\em C-repairs}, are S-repairs that minimize the {\em number} of tuples by which they differ from $D$. The two classes are denoted by $\nit{Rep}(D,\mc{K})$  and  $\nit{Rep}^{c}(D,\mc{K})$, resp. Repairs of $D$ w.r.t. $\mc{K}$ can be characterized as vertex-covers (or hitting sets)\footnote{A hitting-set of a set collection $S$  contains at least one element of each set in $S$.} of the {\em conflict hypergraph}, $\mc{CH}(D,\mc{K})$, whose vertices are the tuples, and hyperedges are minimal sets of tuples that jointly participate in a violation of $\mc{K}$ \cite{icdt07}.


\vspace{-2mm}
\paragraph{\bf Actual Causality and Responsibility.} 
As common in causal settings, where a model may contain  endogenous and exogenous variables, a  DB $D$ is considered to be split in two disjoint sets of  {\em endogenous} and {\em exogenous} tuples: $D = D^n \cup D^x$. The distinction between endogenous and exogenous tuples is application dependent.  For more on this, see Section \ref{sec:disc}, and  \cite[sec. 5]{pearl} for a general discussion.   
\ Sometimes, all tuples will be endogenous, i.e. $D^x = \emptyset$, as in Example \ref{exa:newcrIntro}.  Tuples can be seen as propositional variables, with value ``$1$" if the tuple is in the DB, and ``$0$", otherwise.

  We want to understand the effect of endogenous tuples on the positive answer to a BCQ $\mc{Q}$,  which becomes another, top, propositional variable. Endogenous tuples   are subject to {\em interventions} (think of possibly virtual DB updates; in our context, exploratory deletions), but not the exogenous ones. 

For $\mc{Q}$ a BMQ,  a tuple $t \in D^n$ is  a {\em counterfactual cause} for $\mc{Q}$ being true in $D$ if $D\models \mc{Q}$, but $D\smallsetminus \{t\}  \not \models \mc{Q}$. \  $t \in D^n$ is an {\em actual cause}
if there is $\Gamma \subseteq D^n$,  a {\em contingency set} for $t$, such that $t$ is a counterfactual cause for $\mc{Q}$ in $D\smallsetminus \Gamma$, i.e.  $D \smallsetminus \Gamma \models \mc{Q}$, but $D \smallsetminus (\Gamma \cup \{t\}) \not \models \mc{Q}$ \cite{Meliou2010a}. \  $\nit{Cont}(D, \mc{Q}, t)$ denotes the set of subset-minimal contingency sets for $t$. $\nit{Causes}(D, \mc{Q})$
denotes the set of actual causes.
 The {\em causal responsibility} of $t \in D^n$ is $\rho(t) := 1/(|\Gamma| + 1)$, with $|\Gamma|$ a
minimum-size contingency set for $t$. When $t$ is not an actual cause,  $\rho(t)$ is  $0$.

For  $t \in D^n$, it holds \cite{tocs}:
\ (a) \ $t$ \ is actual cause for $\mc{Q}$ with subset-minimal contingency set $\Gamma$ \ iff \ $D \smallsetminus (\Gamma \cup \{t\}) \subseteq D^n$ and is an  S-repair w.r.t. $\mc{K}(\mc{Q})$. \ (b) \ 
$t$ \ is actual cause with minimum-cardinality contingency set $\Gamma$ \ iff \ $D \smallsetminus (\Gamma \cup \{t\}) \subseteq D^n$  is a C-repair,
in which case, $t$ is a maximum-responsibility actual cause. 
\ It is possible to bring the partition of $D = D^n \cup D^x$ into the notion of repair. In our case, in order to restore consistency, only deletions of endogenous tuples would be accepted \cite{tocs}. This would give rise to  two {\em preferred repair semantics} \cite{CQA}, which we denote by $\nit{Rep}^{n}\!(D,\mc{K})$ and $\nit{Rep}^{\!n,c\!}(D,\mc{K})$.

\vspace{-2mm}
\paragraph{\bf Related Work.}
Along the lines of applications of actual causality to DBs, there have been recent applications to {\em Explainable AI} (XAI),  to explain outcomes from machine learning models. Apart from causality, most  explanations for query
results rely on {\em data provenance} \cite{BunemanKT01}. 

In \cite{Meliou2010a},  a connection between
QA-causality and {\em why-provenance}  \cite{BunemanKT01} was established.  The latter appeals to the notion of {\em lineage} of a query answer and its {\em minimal-witnesses}, which are
subset-minimal subinstances $W \subseteq D$, such that $W \models \mc{Q}$. Also in \cite{Meliou2010a},   causal responsibility
 was applied to the lineage of a query, to provide explanation in terms
 of highest responsible causes. 
 
 A close connection between causal responsibility
and other notions in DBs, e.g. minimal-source-side-effect deletion-problem and
cardinality-based repairs, was established in~\cite{tocs,flairs17}.  The repair core has been used, without that name, in Ontology-Based Data Access, as the repair of an $A$-box w.r.t. a $T$-box \cite{rosati}.

Sufficiency has been recently investigated in data management. In~\cite{ester24a,ester24b}, the authors take the point of view of {\em why-provenance}, and concentrate on Datalog queries, and on the decision version of the problem on the basis of proof-trees. In \cite{bienvenu25}, different aggregations are proposed and investigated to quantitatively identify the most powerful explanations. \ Sufficiency has also been investigated in AI for a long time, also under the notion of {\em abduction},  in DBs \cite{flairs17}, and more recently and actively in the broader area of XAI \cite{ignatiev19}.

\vspace{-2mm}
\section{\ Sufficiency for Query Answering} \label{sec:sufficiency}

\vspace{-2mm}
We start by defining a  measure of the extent by which a tuple is {\em sufficient} to support a positive query answer.\footnote{It was introduced in \cite[chapter 9]{babakThesis}. See \cite{izza} for recent applications in XAI.}

\begin{definition}\label{def:indices} \em
 For  $D=D^x \cup D^n$, a BMQ $\mc{Q}$, with \ $D \models \mc{Q}$, and $t \in D^n$: \
 (a) \  $S \subseteq D^n$ is a {\em  sufficient-set} (SS)  if \ $S \cup D^x \models \mc{Q}$. \
(b) \  $S \subseteq D^n$ is a {\em minimal sufficient-set} (MSS)  if it is a sufficient-set, but
 no proper subset of $S$ is a sufficient-set.\footnote{When $D=D^n$, sufficient-sets  coincide with  {\em minimal-witnesses}  \cite{babakThesis}.}
\ (c) \  The {\em sufficiency-degree} of $t$ is $\sigma(t) := 1/|S|$, with $S$ a minimum-cardinality sufficient-set (mSS) containing $t$.
 If $t$ is not contained in any MSS, $\sigma(t):=0$. Otherwise, $t$ is a {\em sufficient explanation}. If $t$ belongs to every MSS, it is a {\em strong sufficient explanation}.
 \boxtheorem
\end{definition}

\vspace{-3mm}
\begin{example}\label{ex:suffDB} Consider  $D = \{R(c,b), R(a,d), R(b,a),$ $ R(e,f), S(a), S(b),$  $ S(c),$ $ S(d)\}$, with  $D^\nit{x}= \emptyset$; \ and the BCQ $\mc{Q}\!: \exists x \exists y ( S(x) \land R(x, y) \land S(y))$, which is true in $D$. Some proper subinstances suffice to make it true, among them: $S = \{S(c),R(c,b),S(b)\}$,  $S^\prime = \{S(a), R(a,d),$ $ S(d)\}$, and  $S^{\prime\prime} \!= \{S(b), R(b,a),$ $ S(a)\}$. These subinstances are the all and only
MSSs. Here,  they are all also   mSSs. Then, $\sigma(S(b)) = 1/3$.  With $D^\prime = \{R(c,b), R(a,d), R(b,b), R(e,f),$ $ S(a), S(b), S(c)\}$: $S$ and $S^{\prime\prime\prime} \! = \! \{S(b), R(b,b)\}$ are MSSs, and only the second is mSS.  
\boxtheorem \end{example}

\begin{example} \label{ex:exIntro3} (ex. \ref{ex:exIntro2} cont.) \ The sufficiency-degrees are: $\sigma(t_1) =1$, $\sigma(t_2) = \sigma(t_3) = 1/2$, $\sigma(t_4) = \sigma(t_5) = \sigma(t_6) = 1/3$. However, the responsibilities are all $1/3$. \boxtheorem
\end{example}

\vspace{1mm}
\noindent {\bf Notation:} (a) For $D \models \mc{Q}$, $\nit{MSS}(D,\mc{Q})$  denotes the set of  MSSs.  $\nit{MSS}(t)$  denotes the set of MSSs that contain $t$. \\
(b) With $\mc{SH}(D,\mc{Q})$ we denote the {\em sufficiency hypergraph} whose nodes are the tuples of $D$, and hyperedges are the sets of endogenous tuples that together (and possibly accompanied by exogenous tuples) satisfy $\mc{Q}$, that is, all SSs of $\mc{Q}$. \\
(c) We denote by 
$\mathsf{MHS}(\mc{SH})$ (simply $\mathsf{MHS}$ when clear from the context) the set of {\em minimal hitting-sets} (MHSs) of the hyperedges in $\mc{SH}(D,\mc{Q})$ .
\boxtheorem


\vspace{1mm}Notice that $\mc{SH}(D,\mc{Q})$ and $\mc{CH}(D,\mc{K})$ coincide.\footnote{The conflict hypergraph, as usually defined, may include exogenous tuples. This can be easily avoided by redefining it w/o exogenous tuples. On the repair side, one typically considers {\em preferred repairs} that do not touch exogenous tuples.} 
Furthermore, if $\mc{S}$ is the finite family of all SSs, then: (i) Every $S \in \mc{S}$ contains some $S^\prime \in \nit{MSS}(D,\mc{Q})$ as a subset; and (ii) The  family of HSs of $\mc{S}$ coincides with the family of HSs of  $\nit{MSS}(D,\mc{Q})$. Therefore, w.l.o.g.,  
every MHS in $\mathsf{MHS}(\mc{SH})$ can be viewed as a hitting-set of MSSs. 
%
%

\begin{example}\label{ex:abdex3} 
Consider $D = D^n = \{R(a_1,a_4), R(a_1,a_3),$ $ R(a_3,a_3),T(a_1), T(a_2),$ $ T(a_3)\}$, and
  $\mc{Q}\!:  \exists x \exists y(R(x, y) \wedge T(y))$,
which is true in $D$.
 This positive  answer has the
MSSs: \ $S_1=\{R(a_1,a_3),$ $ T(a_3)\}$ and $S_2=\{R(a_3,a_3), T(a_3)\}$.
All tuples have
sufficiency-degree 
$1/2$. 
\boxtheorem\end{example}

\vspace{1mm}
A related notion is that of \emph{necessary tuples} and the necessity-degree~\cite{flairs17}.

\begin{definition} \label{necset} \em
   For $D=D^x \cup D^n$ and a BMQ $\mc{Q}$ with $D \models \mc{Q}$: \
 (a) \ $N \subseteq D^n$ is a {\em  necessary-set}  if \ $ D \smallsetminus N \not \models \mc{Q}$. \
 (b) \  $N \subseteq D^n$ is a {\em minimal necessary-set} (MNS) if it is a necessary-set, but
 no proper subset of $N$ is a necessary-set. \  (c) \  The {\em necessity-degree} of a tuple $t$ is $\eta(t) := 1/|N|$, with $N$ a minimum-cardinality necessary-set (mNS) containing $t$. \ If $t$ is not contained in any  MNS, $\eta(t):=0$. \ Otherwise, we say $t$ is a {\em necessary tuple}. $t$ is {\em strongly necessary} if it belongs to every MNS.
 \boxtheorem
\end{definition}

\begin{proposition} \label{pro:causesuffset}  \em Let $D=D^x \cup D^n$, $\mc{Q}$ be a BMQ
with $D \models \mc{Q}$, and $t \in D^n$. The following hold: \vspace{-1mm}
\begin{enumerate}
    \renewcommand{\theenumi}{\roman{enumi}} 
    \renewcommand{\labelenumi}{(\theenumi)} 
    \item Tuple $t$ \ belongs to some MSS for $\mc{Q}$ iff $t$ is a necessary tuple $\mc{Q}$; \label{itm:causesuffset}
    \item For every $t \in D^n$, if $\sigma(t) > 0$, then  $\eta(t) > 0$. Equivalently, there is no $t \in D^n$ with $\sigma(t) > 0$ and $\eta(t)=0$; \label{itm:sigma-implies-eta}
    \item $S \in \nit{MSS}$ iff it is a minimal hitting-set (MHS) for $\nit{MNS}$; $N \in \nit{MNS}$ iff it is a minimal hitting-set for $\nit{MSS}$. \  \boxtheorem\label{pro:hitting}
\end{enumerate}
\end{proposition}

\vspace{-2mm}
\begin{example} \label{ex:abdex3}  
Consider $D = D^n = \{R(a_1,a_4), R(a_1,a_3),$ $ R(a_3,a_3),T(a_1), T(a_2),$ $ T(a_3)\}$, and
  $\mc{Q}\!:  \exists x \exists y(R(x, y) \wedge T(y))$,
which is true in $D$.
 This positive  answer has the
MSSs: \ $S_1=\{R(a_1,a_3),$ $ T(a_3)\}$ and $S_2=\{R(a_3,a_3), T(a_3)\}$. \  
$S_1$ and $S_2$ are both MHSs -and the only ones- of  $\nit{MNS} = \{\{T(a_3)\}, $ $ \{R(a_1,a_3), R(a_3,a_3)\}\}$. \ Similarly, the latter's elements $\{T(a_3)\}$ and  $\{R(a_1,a_3), R(a_3,a_3)\}$ are both MHSs -and the only ones- of $\nit{MSS} = \{\{R(a_1,a_3),$ $ T(a_3)\}, \{R(a_3,a_3), T(a_3)\}\}$.
\boxtheorem\end{example}

\vspace{1mm}
By exploiting the relation between MSS and MNS, we obtain the following:

\begin{proposition}\label{prop:repairs-mhs} \em Let $\mc{K}$ be the DC associated to $\mc{Q}$. If $N \in \mathsf{MHS}(\mc{SH})$, then $D\smallsetminus N \in \nit{Rep}(D,\mc{K})$.  
Conversely, if $D^\prime \in \nit{Rep}(D,\mc{K})$, $D \smallsetminus D^\prime \in  \mathsf{MHS}(\mc{SH})$.
\ignore{Every MHS $N\in\mathsf{MHS}(\mc{SH})$ is exactly a minimal set of endogenous tuples whose removal from $D$ makes $\mathcal{Q}$ false. Conversely, every minimal set of deletions that yields a repair corresponds to a MHS of $\mc{SH}$.} \boxtheorem
\end{proposition}

In \cite{hu}, the authors establish a dichotomy result for sjf-\!CQs and a certain syntactic class of them related to the free variables in them: For queries in the class, computing a minimum-sufficient set
can be done in polynomial-time (in data), but for those outside the class, it becomes NP-hard.
Our sjf-\!BCQs   (then, w/o free variables) fall inside the class, and consequently, the PTIME result for this class is fully aligned with our results in Section~\ref{sec:explCore}.
They do not consider mSSs containing a given tuple.

From the results in \cite{joaoICML21} about monotone classifiers, it follows that computing a single minimal sufficient explanation can be done in polynomial time. Our case, adapted to that setting, considers  the input feature-vector as formed by the endogenous tuples; and the query and exogenous tuples become part of the monotone binary classifier, which simply evaluates the query.

\ignore{It clear that
causal responsibility of a tuple coincides with its $\nu$ index. In other words, to measure
 causal contribution, responsibility a priori wedded
to the extent by which a tuple is necessary for an answer. Hence, it totally ignores
the effect of the level of sufficiency of a tuple to bring about an answer.}

\vspace{-2mm}
\section{The Repair Core}\label{sec:repair-core}

\vspace{-2mm}
In this section, we will define a useful notion: the repair core. We show its relation to MNSs, MSSs, and the sufficiency- and necessity-degrees. We assume that $D^n \models \mc{Q}$, that is, there is a way to repair $D$ by deleting endogenous tuples.


\begin{definition} \label{def:core} \em Let $\mc{Q}$ a BCQ with associated DC $\mc{K}$, such that $D \models \mc{Q}$.  The  {\em repair core} of $D$ w.r.t. $\mc{K}$ is:

\vspace{-3mm}
\begin{equation}
    \core(D,\mc{K})  \ := \ \bigcap \nit{Rep}^{\!n\!}(D,\mc{K})  = \bigcap_{D' \in  \nit{Rep}^{\!n\!}(D,\mc{K})} \!\!\!\!\!D'. \label{eq:core}
    \end{equation}
 \end{definition}
\vspace{-0.6cm}   \boxtheorem

\vspace{-2mm}
\begin{example}\label{ex:CH}  (ex. \ref{ex:suffDB} cont.)   $D^\prime \not \models \mc{K}$, with $\mc{K}$: $\neg \exists x \exists y ( S(x) \land R(x, y) \land S(y))$. \ The minimal deletion sets are  $\{S(b)\}, \{S(c), R(b,b)\}, \{R(c,b),R(b,b)\}$. \ The only minimum repair is $(D^\prime \smallsetminus \{S(b)\})$.
Thus, according to (\ref{eq:core}),  $
        \core(D',\mc{K}) 
        = \ (D' \smallsetminus \{ S(b) \})$ $ \cap \  (D' \smallsetminus \{ R(c,b), R(b,b) \}) \cap (D' \smallsetminus \{ S(c), R(b,b) \})
        = \ \{ R(a,d), R(e,f), S(a) \}$.
\boxtheorem\end{example}

\vspace{1mm}
Notice that, by definition, $\core$ contains all the exogenous tuples.

\begin{lemma}\label{lem:core-exogenous-decomposition} \em
Let $D = D^n \cup D^x$, $\mc{Q}$ a BCQ, and $\mc{K}$ its associated DC. 
Assume that for every repair $R$ of $D$ w.r.t. $\mc{K}$, $R \cap D^n$ is a repair of $D^n$ w.r.t. $\mc{K}$.
Then, $\core(D,\mc{K}) = D^x \cup \core(D^n,\mc{K})$. \boxtheorem
\end{lemma}

In what follows, we focus on the relationships between $\core$, the MSSs, and the MNSs. 
As a consequence of Proposition~\ref{prop:repairs-mhs}, and the standard identity of intersections of complements with the complement of unions, 
we obtain the following:

\begin{proposition}\label{prop:core-formula} \em
For $D \models \mc{Q}$, and $\mc{K}$ the DC constraint associated to $\mc{Q}$, it holds:
\begin{equation*}\label{eq:core-complement}
   \hspace*{2mm} \core(D,\mc{K}) = 
    \bigcap_{N\in\mathsf{MHS}(\mc{SH})} (D\smallsetminus N) = D \smallsetminus \Big(\bigcup_{N\in\mathsf{MHS}(\mc{SH})} N\Big). 
\end{equation*}

\vspace{-10mm}\boxtheorem
\end{proposition}

\vspace{3mm}
Proposition \ref{prop:core-formula} tells us that a tuple $t\in D$ belongs to $\core$ (equivalently, it is never deleted by any minimal repair) iff it does \emph{not} belong to any MHS.

\begin{example}\label{ex:core-formula} Let $D=\{S(a),S(b),S(c),R(a,b),R(b,c),$ $S(d)\}$, $\mc{Q}\!:\exists x\exists y (S(x)\land R(x,y)\land S(y))$, and   $D^x=\emptyset$.
The MSSs for $\mc{Q}$ are $S_1=\{S(a),R(a,b),S(b)\}$, $S_2=\{S(b),R(b,c),S(c)\}$.
Thus, every tuple in $D\smallsetminus \{S(d)\}$ belongs to some MSS, and has positive sufficiency-degree.
 Here, $\mathrm{MHS}(\mc{SH}) = \{
\{S(b)\}, \{R(a,b),$ $R(b,c)\},$ $ \{R(a,b),S(c)\}, \{S(a),R(b,c)\}\}$.
From Proposition~\ref{prop:core-formula}: $\core(D,$ $\mc{K})=\{ S(d) \}$.
\boxtheorem\end{example}


\begin{proposition}\label{prop:eta-core-iff} \em
Let $D=D^x\cup D^n$ and let $\mc{Q}$ be a BCQ with $D\models\mc{Q}$. 
\begin{enumerate}
    \renewcommand{\theenumi}{\roman{enumi}} 
    \renewcommand{\labelenumi}{(\theenumi)} 
    \item For every $t\in D$, the following are equivalent: (a) $\eta(t)=0$, (b) $t\in\core(D,\mc{K})$, and (c)  $t\notin\bigcup_{N\in\mathsf{MHS}} N$.
    \item Every $t\in \core(D,\mc{K})$ belongs to no MSS. \label{cor:t-core-no-MSS} \boxtheorem
\end{enumerate} 
\end{proposition}




It is important to notice that Propositions~\ref{prop:repairs-mhs} and~\ref{prop:core-formula} hold analogously if $D^x\neq\emptyset$. More precisely: 
\begin{equation}\label{eq:core-mhs-with-ex}
\core(D,\mc{K})
= D \smallsetminus  \!\!\!\!\!\!\!\! \bigcup_{N \in \mathsf{MHS}(\mc{SH})} \!\!\!\!\!\!\!\! N
= D^x \!\;\cup\;\!\!
\Bigl(D^n \smallsetminus \!\!\!\!\!\!\bigcup_{N \in \mathsf{MHS}(\mc{SH})} \!\!\!\!\!\!\!N\Bigr).
\end{equation}
Moreover, if $D^x\neq\emptyset$, Proposition~\ref{prop:eta-core-iff} holds verbatim: MNSs are, by definition, subsets of  $D^n$. Hence, no exogenous tuple can belong to any MNS. This observation, although conceptually simple, is  relevant: The necessity-score of an exogenous tuple is trivially zero, because exogenous tuples are not eligible for deletion, and thus, cannot be ``necessary'' for falsifying the query: The necessity score treats exogenous tuples as structurally irrelevant for the purposes of repair-based explanations. \ Furthermore, under our assumptions, exogenous tuples do not occur in MSSs, and, therefore, we also get $\sigma(t)=0$.
The following proposition establishes the relationship between the sufficiency-score and the core.



\begin{example}  (ex.~\ref{ex:core-formula} cont.)\label{ex:core-exogenous}
Now assume that $D^{x}=\{S(b)\}$ and $D^{n}=D\smallsetminus D^{x}$. Recall that MSSs and MNSs are subsets of $D^{n}$. \ The MSSs are  $S_1=\{S(a),$ $R(a,b)\}$ and  $S_2=\{R(b,c),S(c)\}$; \ignore{,
since $S_i\cup D^{x}\models\mc{Q}$ and both sets are minimal with this property.}
and the MHS for them are: \ $
\{R(a,b),R(b,c)\},$ $ \{R(a,b),S(c)\}, \{S(a),$ $R(b,c)\}$ and $ \{S(a),S(c)\}$.

Hence, according to (\ref{eq:core-mhs-with-ex}), $\core(D,\mc{K}) = D\smallsetminus \{S(a),$ $S(c),R(a,b),$ $R(b,c)\}
= \{S(b),S(d)\}$.
Notice  that since $S(b)$ is exogenous, it belongs to $\core$, but not to
any MSS or MNS. 
\boxtheorem\end{example}

\vspace{-2mm}
\section{Sufficient  Explanations via the  Core}\label{sec:explCore}

\vspace{-1mm}
In Section \ref{sec:pre}, we described the connection between necessary explanations and DB repairs.
\ In this section, we will establish useful connections between sufficient explanations and DB repairs;  and will use the repair core for computing sufficient explanations.
To simplify the presentation, we assume DBs have no exogenous tuples. The results in this section can be easily extended to the general case. We will expand on this at the end of the section (see Remark~\ref{obs:why-no-exogenous-ok}).

The number of repairs can be exponential in the size of the $D$ \cite{CQA}. Then, computing $\core$ by first computing all repairs may take exponential time. The next result will allow us to efficiently compute  $\core$. \ignore{in polynomial time in the size of $D$.}

\ignore{\red{Here, we have and instance $D$ a BCQ $\mc{Q}\!: \ \exists \bar{x}\, (P_1(\bar{x}) $ $\land \ldots \land  P_k(\bar{x}))$, such that $D \models \mc{Q}$, and  the associated DC $\mc{K}\!: \  \neg \exists \bar{x}\ (P_1(\bar{x}) \land \ldots \land  P_k(\bar{x}))$.  With $P^D$ we denote the extension of predicate  $P$ in  $D$. }
 \nina{hace falta repetir la forma de $\mc{Q}$ en el statement del lemma?}
 }

\vspace{-1mm}
\begin{lemma}\label{claim:1_core_rewriting} \em For an instance $D = D^x \cup D^n$ and a sjf-\!BCQ \ $\mc{Q}\!: \exists \bar{x}\, (P_1(\bar{x}) \land \ldots \land  P_k(\bar{x}))$ with associated DC $\mc{K}$, and  \
    every $i \in \{ 1, \ldots, k\}$: \ If
        $R_i^D \ := \ \{ t \in P_i^D \ \mid \ \mbox{ for every } j \in [1,k]\smallsetminus \{i\}, \mbox{ then, there is } t_j \in P_j^D,   \mbox{ such that }
         \{t \} \ \cup \ \bigcup_{j\neq i}\{ t_j\} \models P_1(\bar{x}) \land \ldots \land  P_k(\bar{x}) \}$. Thus, \  $\core(D,\mc{K}) \ = \ \bigcap_{1 \leq i \leq k } (D \smallsetminus R_i^D)$. \boxtheorem
\end{lemma}



\vspace{-1mm}
Lemma~\ref{claim:1_core_rewriting} gives us an algorithm for $\core$ computation that avoids building or inspecting every possible repair; and deterministically build  $\core$ in polynomial time in the size of $D$. Actually, we can build the sets $R_i^D$ in $|D|^{|\mc{Q}|}=|D|^k$-time, as follows:  (1)
 Build $R^D_1$ by picking  a tuple $t \in P^D_1$.  (2) Try all possible combinations of tuples in $P_2^D, \ldots, P_k^D$, checking if, together, they satisfy $\mc{Q}$. This gives at most $|D|^{k-1}$ options for each tuple $t$ in $P_1^D$.  (3) We obtain an upper bound of $|D|^k$ for the computation of each $R_i^D$.

\begin{corollary}\label{cor:core_complexity_DC} \em
    For a (fixed) sjf-\!BCQ $\mc{Q}$ of size $k$, and an instance $D$,  $\core(D,\mc{K})$ can be computed in $\mathcal{O}(|D|^k)$, that is, in PTIME in the size of  $D$.  \boxtheorem
\end{corollary}

\vspace{-2mm}\begin{remark}
    Lemma~\ref{claim:1_core_rewriting} is not true for non-sjf-\!BCQ. Consider $\mc{Q}: \exists x \exists y(R(x,y) \wedge R(y,y))$, and $D=\{R(a,a),R(b,a)\}$. \ $\core(D, \mc{K})=\{R(b,a)\}$ since this is the only repair. The set $R_1^D$, as defined in the lemma, would be $\{R(a,a),R(b,a)\}$, which is $D$. Hence, the equality stated in the lemma does not hold. 
\end{remark}

%


\begin{example}\label{ex:enumeration-running-ex}  
Consider $D = D^n = \{R(a_1,a_3), R(a_1,a_4),R(a_3,a_3),$ $S(a_1), T(a_2)$, $T(a_3),$ $T(a_4)\}$, and
  $\mc{Q}\!:  \exists x \exists y(S(x) \wedge R(x, y) \wedge T(y))$.
For the given $D$ and $\mc{Q}$, here $k=3$ and $|D|=7$, with atom positions $P_1=S$, $P_2=R$ and $P_3=T$.

To construct $R_1^{D}$ (tuples of $S$ that participate in a witness of $\mc Q$), for each $t\in P_1^{D}=S$, we try all combinations of one tuple from $P_2^{D}=R$ and one tuple from $P_3^{D}=T$, and check whether together they satisfy $\mc Q$:

\noindent (i) \
$S(a_1)$ participates (with $R(a_1,a_3),T(a_3)$, and also with $R(a_1,a_4),T(a_4)$). \
Hence,
\(
R_1^{D}=\{S(a_1)\}.
\)

\noindent (ii) \
We repeat the procedure for $R_2^{D}$:
$R(a_1,a_3)$ participates (with $S(a_1),T(a_3)$). \
$R(a_1,a_4)$ participates (with $S(a_1),T(a_4)$). \
$R(a_3,a_3)$ does not participate (since $S(a_3)\notin D$). \
Thus,
\(
R_2^{D}=\{R(a_1,a_3),R(a_1,a_4)\}.
\)

\noindent (iii) \
Finally, we build $R_3^{D}$:
$T(a_2)$ does not participate (no tuple $R(\_,a_2)$). \
$T(a_3)$ participates (with $S(a_1),R(a_1,a_3)$). \
$T(a_4)$ participates (with $S(a_1),R(a_1,a_4)$). \
Hence,
\(
R_3^{D}=\{T(a_3),T(a_4)\}.
\)

\noindent We perform all possible tuple combinations in the other two atom positions,
with at most $|D|^{k-1}=7^2$ checks per $t$, and at most $|D|^k=7^3$
in total, for each $R_i^{D}$.


 Note that the minimal deletion sets are  $\{S(a_1)\}$, $\{T(a_3)$, $R(a_1,a_4)\}$, $\{T(a_4)$, $R(a_1,a_3)\}$, $\{R(a_1,a_4)$, $R(a_1,a_3)\}$, and $\{T(a_3), T(a_4)\}$. \ The only minimum repair is $(D \smallsetminus \{S(a_1)\}$.
Here, $k=3$, and $R_1^{D} = \{ S(a_1) \}$, $R_2^{D}= \{ R(a_1,a_3), R(a_1,a_4) \}$, and $R_3^{D}= \{ T(a_3), T(a_4) \}$. \ According to Lemma~\ref{claim:1_core_rewriting}:
    \noindent $\core(D) = \bigcap_{i=1,2,3} (D \smallsetminus R_i^{D})
        = (D \smallsetminus \{ S(a_1) \}) \cap (D \smallsetminus \{ R(a_1,a_3), R(a_1, a_4) \})\cap \ (D \smallsetminus \{ T(a_3), T(a_4) \})
        = \{ R(a_3,a_3), T(a_2) \}$.
\ This coincides with what we can obtain directly from the definition with the minimum repairs. 
\boxtheorem\end{example}

\vspace{-1mm}
\begin{proposition}\label{prop:t_suf_exp} \em
    Consider DB $D$, a sjf-\!BCQ $\mc{Q}$, such that $D \models \mc{Q}$, $t \in D$, and associated DC $\mc{K}$. \
    Let $D'$ be a repair of $D$ \wrt $\mc{K}$. It holds: \ $t \in D\smallsetminus D'$ iff
    $(D' \smallsetminus \core(D,\mc{K})) \cup \{t\}$ is a sufficient-set (SS) for $\mc{Q}$ \ (not necessarily minimal or minimum).
    \boxtheorem 
\end{proposition}





\vspace{-3mm}
\begin{example}\label{ex:prop1} (ex. \ref{ex:enumeration-running-ex} cont.) 
    $D^\prime := D \smallsetminus \{ S(a_1) \}$ is a repair of $D$. Furthermore, $D^\prime \smallsetminus \core(D) = \{ R(a_1,a_3), R(a_1,a_4), T(a_3), T(a_4) \}$, and $\big( D^\prime \smallsetminus \core(D) \big) \not\models \mc{Q}$. Moreover, $D \smallsetminus D^\prime = \{ S(a_1) \}$.
    Then, $\{ S(a_1)\} \cup (D^\prime \smallsetminus \core(D))$ is a SS for $\mc{Q}$. 

    Conversely, assume there is $t$ such that $(D^\prime\smallsetminus \core(D)) \cup \{ t\}$ is a SS for $\mc{Q}$. In particular, we have that $ D \smallsetminus ( D^\prime\smallsetminus \core(D) ) = \{ S(a_1), R(a_3,a_3), T(a_2) \}$, where $S(a_1)$ is the only tuple in $D \smallsetminus D^\prime$.
    Moreover, it is easy to see that, for every $t \in \{ R(a_3,a_3), T(a_2) \}$, $\{t\} \cup \{ R(a_1,a_3), R(a_1,a_4), T(a_3), T(a_4) \} \not\models \mc{Q}$,  but $\{ S(a_1) \} \cup \{ R(a_1,a_3), R(a_1,a_4), T(a_3), T(a_4) \} \models \mc{Q}$. 

\boxtheorem\end{example}

\vspace{1mm}
We will see next that the second condition in Proposition \ref{prop:t_suf_exp} can be replaced by ``contains a MSS for $\mc{Q}$ that contains $t$". Actually, Proposition \ref{prop:t_suf_exp_alg} below provides a chase-like procedure to construct such a MSS contained in $(D'\smallsetminus \core(D)) \cup \set{t}$. \
The algorithm uses the characterization of $\core$ given in Lemma~\ref{claim:1_core_rewriting}. The
proposition tells us that any $t \in D\smallsetminus D^\prime$,
with $D^\prime$ a repair, is a good starting point for the procedure. The latter will terminate since we know for sure, by Proposition \ref{prop:t_suf_exp}, that we can obtain a sufficient-set from any such tuple. 

Before proceeding, we first introduce some notation. \ 
For an instance $D$ and query $\mc{Q}$ as in Lemma \ref{claim:1_core_rewriting}, let $s \in P_i^D, t\in P_j^D$ be two distinct tuples in two different relations of $D$. We denote with  $s\vert_{t}$ the subtuple of $s$ obtained by restricting $s$ to the values of those variables in common with $t$ according to $\mc{Q}$. For example, for $\mc{Q}\!: \bar{\exists}(S(x,y,z) \wedge T(x,z,u,v))$, and $s=S(0,1,2)$ and $t=T(0, 1, 2, 3)$: $s\vert_{t} = (0, 2)$.

\ignore{
$s=(x:0, y:1, z:2)$ and $t=(x:0, z:1, u:2, v:3)$, then $s\vert_{t} = (x:0, z:2)$.}

\vspace{-4mm}
\begin{proposition}  \label{prop:t_suf_exp_alg} \em
For an instance $D$, a sjf-\!BCQ \ $\mc{Q}\!: \exists \bar{x}\, (P_1(\bar{x}) \land \ldots \land  P_k(\bar{x}))$, $D^\prime$ a repair of $D$ w.r.t. $\mc{K}$, and   $t \in (D\smallsetminus D') \cap P_1^D$ (w.l.o.g.),  it holds:

\noindent (a) There is a set in $\nit{MSS}(t)$ contained in $(D' \smallsetminus \core(D,\mc{K})) \cup \{t\}$. 

\vspace{1mm} \noindent (b) A MSS $S$ as in (a) of size at most $k$ can be computed \ignore{the form $\{t_1, \ldots, t_k\}$}\ignore{$(D' \smallsetminus \core) \cup \{t\}$} by the following procedure: \vspace{-4mm}
\begin{enumerate}
\item Starting with $t=t_1 \in P^D_1$, take $t_2 \in P_2^D$ with $t_2 \in (D' \smallsetminus \core(D, \mc{K}))$ and $t_2\vert_{t_1}= t_1\vert_{t_2}$.

\item Take $t_3 \in P_3^D$ with $t_3 \in (D' \smallsetminus \core(D,\mc{K}))$, \ $t_3 \vert_{t_1}= t_1\vert_{t_3}$, and $t_3 \vert_{t_2}= t_2\vert_{t_3}$.

\item Repeat the inductive procedure until obtaining $S = \{t_1, \ldots, t_k\}$, with $k$ the number of atoms in the query. \ 
Since we are assuming that $\mc{Q}$ is sjf, the procedure will stop after reaching $k$ non-trivial steps.\boxtheorem
\end{enumerate}
\end{proposition}





\vspace{-5mm}
\begin{example}\label{ex:prop2}  (ex. \ref{ex:enumeration-running-ex} cont.) 
    Consider $D^\prime = D \smallsetminus \{ S(a_1) \}$ a repair of $D$, thus $D \smallsetminus D^\prime = \{ S(a_1) \}$. Recall that $D^\prime \smallsetminus \core(D) = \{ R(a_1,a_3), R(a_1,a_4), T(a_3), T(a_4) \}$. 
    Consider the only tuple $t \in (D \smallsetminus D^\prime)\cap P_1^{D}$, that is $t=S(a_1)$.
    By Proposition~\ref{prop:t_suf_exp_alg}, we obtain a MSS that contains $t$ as follows:
\ Take $t_2 \in P_2^{D}=R$ with $t_2 \in (D^\prime \smallsetminus \core(D))$ and $t_2\vert_{t}= t\vert_{t_2}$. We have two options, $t_2=R(a_1,a_3)$ or $t_2=R(a_1,a_4)$; take $t_2=R(a_1,a_3)$.
\  Now, take $t_3 \in P_3^D=T$ with $t_3 \in (D^\prime \smallsetminus \core(D))$. Moreover, $t_3 \vert_{t}= t\vert_{t_3}$--which in this case does not add anything since $t_3 \vert_{t}$ is empty--and $t_3 \vert_{t_2}= t_2\vert_{t_3}$, thus $t_3=T(a_3)$ is the only option.
    Since $k=3$, we have finished the procedure, then, $S = \{S(a_1), R(a_1,a_3), T(a_3)\}$ is a MSS that contains $t=S(a_1)$. \  Notice that, since $S$ is both a MSS and a mSS,  $\sigma(t) = 1/|\mathcal{S}|= 1/3$. 
\boxtheorem\end{example}

The previous results in this section allow us to obtain a sufficient-set, and then, a MSS, by simply using the core, without having to either inspect every possible repair, nor checking if every possible subset of tuples derives the query. \ Next, we refer to the procedure described as {\bf Algorithm 1} below; summarizing the results obtained in this section, in Theorem \ref{theo:main} and its Corollary.

\vspace{-4mm}
\begin{algorithm}[H]
\caption{Procedure to compute a MSS}
\label{alg:mss_construction} {\scriptsize
\begin{algorithmic}[1]
    \State \textbf{Input:} $D$, $\mc{K}$, $\mc{Q}$.
    \State \textbf{Output:} A MSS for $D$ \wrt $\mc{Q}$.

    \Statex \textit{\textbf{1. Compute $\core(D,\mc{K})$.}}
    \State By Corollary~\ref{cor:core_complexity_DC}, the $\core$ can be computed in polynomial time (in data complexity).

    \vspace{-2mm}
    \Statex
    \Statex \textit{\textbf{2. Find a suitable tuple $t$}}

    \Statex \textbf{Option 1:}
    \State Choose $t$ from $D \smallsetminus \core(D)$.
    \Comment{\textit{Any tuple not in $\core$ is, by definition, not present in \textbf{at least one repair}.}}

    \Statex \textbf{Option 2 (Non-deterministic): find a repair $D^\prime$}
    \State Compute a repair $D'$ of $D$ non-deterministically.
    \State Let $t \in D \smallsetminus D'$. This repair $D'$ can be used in the next step.

\vspace{-2mm}
    \Statex
    \Statex \textit{\textbf{3. Construct a MSS:}}
    \State \hspace{\algorithmicindent} Using algorithm in Proposition~\ref{prop:t_suf_exp_alg}, compute a \textbf{MSS} for $D$ \wrt $\mc{Q}$.

    \Comment{\textit{Input: $\core$, a tuple $t \in D \smallsetminus \core(D)$ (resp.\ $t \in D \smallsetminus D'$ for Option 2). This finds a MSS that includes $t$. This step runs in polynomial time.}}
\end{algorithmic} }
\textbf{Note:} Since $(D\smallsetminus \core(D)) \cup \{t \} \supseteq (D^\prime\smallsetminus \core(D)) \cup \{t \}$, a MSS in the latter is contained in the former, thus we do not need to explicitly compute the repair $D'$ to run the algorithm.
\end{algorithm}

\vspace{-8mm}
\begin{theorem} \label{theo:main} \em
Given an instance $D$ and a (fixed) sjf-\!BCQ $\mc{Q}$ with $D \models \mc{Q}$, and the associated DC $\mc{K}$, the procedure in {\bf Algorithm~\ref{alg:mss_construction}} computes a MSS in polynomial-time in the size of $D$.
\boxtheorem
\end{theorem}

\vspace{-1mm}
Theorem~\ref{theo:main} is constructive: It \textit{finds} a tuple, and builds an
MSS for $\mc{Q}$. 
Now, if we are given 
a tuple $t\in D$, and $t \in \core(D,\mc{K})$,  we showed in Proposition~\ref{prop:eta-core-iff}(\ref{cor:t-core-no-MSS}) 
that $t$ belongs to no 
MSS.
Otherwise, $t$ occurs in some MSS, and thus we can feed $t$ to our algorithm. Since $\mc{Q}$ is a sjf-\!BCQ,  from Theorem~\ref{theo:main} and
Proposition~\ref{prop:t_suf_exp_alg}, it follows that the constructed MSS will contain tuples for different predicates, yielding a mSS
containing $t$.

\begin{corollary}\label{cor:sjf+} \em Let $\mc{Q}$ be a (fixed) sjf-\!BCQ, $D$ be an instance s.t. $D \models \mc{Q}$, and $t \in D$. If $t\not\in\core(D,\mc{K})$, a mSS containing $t$ can be computed in PTIME in the size of $D$, and the same applies to the sufficiency-score for $t$.
Otherwise, no such set exists and the
sufficiency-score of $t$ is $0$.\boxtheorem
\end{corollary}


The following example shows that the corollary cannot be obtained from Theorem \ref{theo:main} when the query is not SJF.

\vspace{-1mm}

\begin{example} \label{ex:counter}  Consider  $D = \{R(a,b), R(b,b), R(b,c), $$
             R(a,a), S(a,b), S(b,c),$ \linebreak $ S(a,a)\}$, with $D^x = \emptyset$; and  $\mc{Q}\!: \exists x \exists y \exists z \ (R(x, y) \land R(y, z) \land S(x, y))$, which is true in $D$.

    \ Here, tuple $t = R(a,a)$ belongs to a sufficient set for $\mc{Q}$. It holds: (a) $S_1 = \{ R(a,a), S(a,a) \}$ is a mSS; and (b) $S_2 = \{ R(a,b), R(b,c), S(a,b) \}$ is a MSS, but not a mSS. \
    It is easy to see that all tuples except $S(b,c)$ appear in some relation-specific sufficient-set $R_i^D$, for $i=1,2,3$, corresponding to the atoms $R(x,y)$, $R(y,z)$, and $S(x,y)$, respectively: \
    Indeed,
      $\{ R(a,a), R(a,\_), S(a,a) \}$ and
      $\{ R(a,b), R(b,\_),$ $ S(a,b) \}$
    are all the sufficient-sets for the query $\mc{Q}$, where $R(a,\_)$ could be either $R(a,b)$ or $R(a,a)$; similarly for $R(b,\_)$.
    Therefore, every tuple in $D$ belongs to some $R_i^D$, except for $S(b,c)$. \ Hence,  $\core= \bigcap_{i=1,2,3} (D \smallsetminus R_i^D)$ is precisely $\{S(b,c)\}$.

    Now, consider the following repair of $D$: \
    $D^\prime = \{R(a,b), R(b,b), R(b,c),  S(a,a),$ $ S(b,c)\}$.
\ Let $t = R(a,a) \in D\smallsetminus D'$.
    If we consider $(D^\prime \smallsetminus \core) \cup \{t\}$, we get: \
    $A := (D^\prime \smallsetminus \core) \cup \{t\} = \{ R(a,b), R(b,b), R(b,c), S(a,a), R(a,a) \}$,
    which, as we have shown, is a SS.

    Observe that, in the sufficient set $A$, we can obtain a MSS that includes $t$, such as $S_2$. Moreover, the algorithm could also obtain a  mSS, namely $S_1$.
    Thus, depending on the chosen repair and the behavior of the algorithm, it is possible to obtain a MSS that includes $t$, but is not minimum. 
\boxtheorem\end{example}

\vspace{-1mm}
\begin{remark}\label{obs:why-no-exogenous-ok} 
All the results in this section remain valid in the presence of exogenous tuples, with some modifications.
    In Lemma~\ref{claim:1_core_rewriting}, each set $R_i^D$ should be defined over endogenous tuples only, that is, $R_i^D \subseteq P_i^D \cap D^n$.
    Under this assumption, $D^x \cap R_i^D =\emptyset$, and thus, $D^x \subseteq \core(D,\mc{K})$, as expected.
    Proposition~\ref{prop:t_suf_exp} is intended to apply to endogenous tuples only.
    Indeed, if $t \in D^x$, then $t$ belongs to every repair of $D$, and hence, cannot satisfy $t \in D \smallsetminus D'$ for any repair $D'$. Accordingly, the equivalence in the proposition holds under the implicit assumption that $t \in D^n$.
    Proposition~\ref{prop:t_suf_exp_alg} is unaffected by the presence of exogenous
    tuples. Since $D^x \subseteq \core(D,\mc{K})$, all tuples selected from
    $D' \smallsetminus \core(D,\mc{K})$ are endogenous. Thus, the construction of MSS in the proposition operates entirely over $D^n$.
\end{remark}

\vspace{-1mm}
\begin{remark}   It is easy to see that all results in this section remain valid when the sjf-\!BCQ contains predicates whose complete extensions in the DB at hand are formed by exogenous tuples. At each step of our iterative algorithm, those predicates and instances can be ignored. 
\end{remark}

\vspace{-2mm}
\section{ASP-Based Specifications of MSSs}\label{sec:asp}

\vspace{-2mm}
We wonder if we can specify and compute via {\em answer-set programs} (ASPs), also called {\em logic programs with stable-model semantics} \cite{brewka}, the necessary- and sufficient-sets for a conjunctive query on a DB. The former can be computed using ASP appealing to the actual-causality connection.\footnote{Actually, the intersection of the models of this ASP-program contains the core, and can be obtained via skeptical query answering on the program.} Since this can be done with the results in  \cite{kais}, including the specification of minimum-size necessary-sets and computation of {\em responsibility},  we  concentrate on sufficient explanations.

It is worth mentioning that, appealing to Proposition \ref{pro:causesuffset}, one could add an extra layer of ASP-program on top of the ASP that computes necessary explanations; a layer that computes hitting sets, something that can be done with ASP\ignore{(for this and much more, see \cite{meel})}. However, this approach also adds an extra layer of complexity, going up one  level in the polynomial-hierarchy in data complexity \cite{dantsin}; and also an off-line processing of the results of the first program. Here, instead, we show a general direct approach.

The ASP $\Pi(D,\mc{Q})$ that specifies SSs contains the following elements:\footnote{We will use the ``classic" DLV notation \cite{leoneCompLogic}, and that of DLV-Complex \cite{calimeri09} for set building and counting. A ``standardized" syntax can be found in \cite{schaub}.} 

\vspace{1mm}
\noindent {\bf 1. \ Facts (atoms) and Annotated TIDs:} 
\ Facts of the form $R(t_{\bar{c}},\bar{c},\mbf{n})$, $R(t_{\bar{c}^\prime},\bar{c}^\prime,\mbf{x})$, for all tuples $R(\bar{c})$ $ \in D^n$, $R(\bar{c}^\prime) \in D^x$.
\ Here, $t_{\bar{c}}, t_{\bar{c}^\prime}$ are constants for universal tuple-identifiers (tids), and $\mbf{n},\mbf{x}$ annotation constants for ``endogenous", resp. ``exogenous". tids are not essential, but highly convenient. In the rest, we can keep working mostly with tids, as supported by a predicate defined by the following rule that generated as new facts all the combinations of tids and associated annotations for later use:\vspace{-2mm}
\begin{eqnarray}
\nit{tidAn}(t,a)&\leftarrow& R(t,\bar{c},a);  \mbox{ with } R(\bar{c}) \in D.\label{eq:annot}
\end{eqnarray}

\vspace{-1mm}
\noindent {\bf 2. Rules for Satisfying Joins:} \ For the BCQ  $\mc{Q}\!: \bar{\exists}(P_1(\bar{s}_1) \wedge \cdots \wedge P_m(\bar{s}_m))$:\vspace{-1mm}
\begin{eqnarray}
\nit{satCb}(t_1,\ldots,t_m) &\leftarrow& P_1(t_1,\bar{s}_1,a_1), \ldots, P_m(t_m,\bar{s}_m,a_m). \label{eq:pick}
\end{eqnarray}

\vspace{-2mm}Here, all the arguments are variables (except for constants  appearing in the $\bar{s}_i$), the last ones for annotations. The head predicate in (\ref{eq:pick}) stands for a ``satisfying combination" (of tids). That predicate will collect all the combinations of tids (then, implicitly of associated tuples) that make the query true; and nothing more, by the minimality of stable models. 

If the query has self-joins, a combination may not correspond to a MSS when transformed into a set (with rules 4. below). Otherwise, they will become MSSs.
If we are interested only in mSSs, this is not a problem, since with 5. below, we can keep only those.
\ For non-sjf-\!CQs, in order to obtain only MSSs, we can add conditions in the rule body of (\ref{eq:pick}) that depend on the query. For illustration, assume that $P_1$ and $P_2$ are the same predicate, and the rest are all different. We replace (\ref{eq:pick}) by the following rules (that depend only on $\mc{Q}$): \vspace{-1mm}
{\scriptsize \begin{eqnarray*}
\nit{OnlyP}_{\!\!1}(t_1,\ldots,a_m)& \leftarrow & P_1(t_1,\bar{s}_1,a_1), P_1(t_1,\bar{s}_1,a_1), P_3(t_3,\bar{s}_3,a_3), \ldots, P_m(t_m,\bar{s}_m,a_m).\\
\nit{satCb}(t_1,t_1,t_3,\ldots,t_m) &\leftarrow&\nit{OnlyP}_{\!\!1}(t_1,\ldots,a_m).\\
\nit{satCb}(t_1,\ldots,t_m) &\leftarrow& P_1(t_1,\dots), P_2(t_2,\ldots),\ldots, P_m(t_m,\ldots), \nit{not} \nit{OnlyP}_{\!\!1}(t_1,\ldots,a_m).
\end{eqnarray*}}

\vspace{-2mm}
The first rule captures the case where the query is satisfied through the join of two identical atoms. The next one, is as before. The last rule,  the case where the join is satisfied, but not  using two identical ground body-atoms. 

\vspace{1mm}
\noindent {\bf 3. One Combination per Model:} \ Up to now, every stable model will contain all the join-combinations that satisfy the query. Since we want to select combinations with a good property, we create one model per combination, by means the ``choice operator"\cite{zaniolo}:\vspace{-2mm}
\begin{eqnarray}
\nit{preSS}(t_1,t_2,t_3,t_4)&\leftarrow&\nit{satCb}(t_1,t_2,t_3,t_4), \msf{choice}(t_1,t_2,t_3,t_4).\label{eq:choice}
\end{eqnarray}

\vspace{-2mm}
This has the effect of choosing one element of $\nit{satCb}$, and putting it into predicate  ``pre-sufficient-set". Stable models will have each only one join-combination.\footnote{The ``operational" predicate $\msf{choice}$ can be eliminated, defining it in a general manner by means of a few regular rules that involve non-stratified negation \cite{zaniolo}.}  In 5. below, we will get rid of those models with undesirable combinations.

\vspace{2mm}
\noindent {\bf 4. Building SSs:} \ The $\nit{preSS}$ in each stable model is recursively transformed into an ``endogenous" sufficient-set, by means of  the set-construction and union operators \cite{calimeri09}: 
\vspace{-4mm}
{\scriptsize \begin{eqnarray*}
\nit{SS}(\{t_1\}) &\leftarrow& \nit{preSS}(t_1,t_2,t_3,t_4), \nit{tidAn}(t_1,\msfbf{n}).\\
\nit{SS}(\nit{\#union}(s,\{t_2\})) &\leftarrow& \nit{SS}(s),  \nit{preSS}(t_1,t_2,t_3,t_4), \nit{tidAn}(t_2,\msfbf{n}).\\
\nit{SS}(\nit{\#union}(s,\{t_3\})) &\leftarrow& \nit{SS}(s),  \nit{preSS}(t_1,t_2,t_3,t_4), \nit{tidAn}(t_3,\msfbf{n}).\\
\nit{SS}(\nit{\#union}(s,\{t_4\})) &\leftarrow& \nit{SS}(s),  \nit{preSS}(t_1,t_2,t_3,t_4), \nit{tidAn}(t_4,\msfbf{n}).
\end{eqnarray*}}

\vspace{-2mm}Due to the minimality of stable models, we obtain all and only minimal SSs.

\vspace{1mm}
\noindent {\bf 5. Minimizing Sufficiency-Sets:} \ By means of a {\em weak constraint} \cite{leoneCompLogic}, we keep only those stable models whose  associated SSs have a minimum size  (another source on unstratified negation): \vspace{-4mm}
\begin{eqnarray}
&\Leftarrow& \ \nit{SS}(t,\msfbf{n}).\label{eq:wc}
\end{eqnarray}

\vspace{-2mm}
Now the stable models are in one-to-one correspondence with the mSSs.\footnote{As a {\em hard constraint}, written ``$\leftarrow \nit{SS}(t,\msfbf{n}).$", it would discard all models where SSs contain endogenous tuples. As a weak constraint, it minimizes their number.}  

\vspace{1mm}
\noindent {\bf 6. Computing Sufficiency-Degrees:} \
Finally, we can compute the (shared) cardinality of mSSs.  We use the counting  operator \cite{calimeri09}: \vspace{-2mm}
\begin{eqnarray*}
\nit{cardSS}(s,n) &\leftarrow& \nit{SS}(s), \msf{\#int}(n),\msf{\#count}\{t: \msf{\#member}(t,s)\} = n.\\
\nit{invSD}(t,n) &\leftarrow& \nit{SS}(s), \msf{\#member}(t,s), \nit{cardSS}(s,n).,
\end{eqnarray*}

\vspace{-2mm}\noindent with the last rule computing the  inverse of the  sufficiency-degree of an endogenous tuple that  belong to a mSS.

Without the weak-constraint (\ref{eq:wc}),  our ASPs computes all mSSs and MSSs for arbitrary CQs, and all those (stable models associated to those) sets can be queried simultaneously, under the {\em skeptical} or the {\em brave} semantics. For example, we can skeptically obtain those tuples that belong to {\em all} MSSs or  {\em all} mSSs (strong sufficient tuples); and bravely, those that belong to {\em some} of them (weak sufficient tuples). To obtain the (inverse of the) sufficiency-degree of a particular tuple, $\mbf{t}$, we can bravely pose the query: \ $\mbox{:-?} \  \nit{invSD}(\mbf{t},X)$. 

\vspace{-1mm}
\begin{example}\label{ex:counterExo} Modifying Ex. \ref{ex:counter}, consider  $D = \{R(a,b),$ $ R(b,b), R(b,c), R(a,a),$ $ S(a,b), S(b,c),S(a,a), T(a,a)\}$, with $D^x = \{T(a,a)\}$; and  $\mc{Q}\!: \exists x \exists y \exists z \ (R(x, y) \wedge R(y, z)$ $ \wedge S(x, y) \wedge T(x,y))$, which is true in $D$.  Again, $S_1 =$  $ \{ R(a,a), S(a,a) \}$ is a mSS; but  $S_2 = \{ R(a,a),$ $ R(a,b), S(a,a) \}$ is a SS, but neither a mSS nor a MSS (it is contained in $S_1$).

Rules 3.-6. above are quite generic. The relevant specific facts and rules for this example are: \ Facts: \ $R(\msfbf{t1},a,b,\msfbf{n}),$  $ R(\msfbf{t2},b,b,\mbf{n}),$ $ R(\msfbf{t3},b,c,\mbf{n}),$ $ R(\msfbf{t4},a,a,\mbf{n}),$ $ S(\msfbf{t5},a,b,\mbf{n}),$ $ S(\msfbf{t6},b,c,\mbf{n}),$ $S(\msfbf{t7},a,a,\mbf{n}),$ $ T(\msfbf{t8},a,a, \mbf{x})$, all ground atoms.

\vspace{1mm}
\noindent Rules: \ $\nit{tidAn}(t,a)  \leftarrow R(t,x,y,a)$ \  (similarly for $S, T$), plus: \vspace{-2mm}
{\scriptsize \begin{eqnarray*}
\nit{OnlyR}(t_1,t_1,t_2,t_4) &\leftarrow& R(t_1,x,y,a_1), R(t_1,y,z,a_2),S(t_3,x,y,a_3), T(t_4,x,y,a_4).\\
\nit{satCb}(t_1,t_1,t_3,t_4) &\leftarrow& \nit{OnlyR}(t_1,t_1,t_3,t_4).\\
\nit{satCb}(t_1,t_2,t_3,t_4) &\leftarrow& R(t_1,x,y,a_1), R(t_2,y,z,a_2),  S(t_3,x,y,a_3), T(t_4,x,y,a_4),\\ &&~~\nit{not} \nit{OnlyR}(t_1,t_1,t_2,t_4).
\end{eqnarray*}}

\vspace{-3mm}In this case, the program will return only the stable model containing $S_1$, even without (\ref{eq:wc}). \boxtheorem
\end{example}

 \vspace{-2mm}
 \paragraph{\bf Generic Solutions:} General ASPs for specifying solutions to {\em abductive problems} of the form $\nit{AP} = \langle H, M, O\rangle$ were proposed \cite{eiterLeone}, where  $H$ is a set of possible {\em hypothesis} ({\em abducible atoms}), $M$ is a system's {\em model} written as  some sort of logic program, and $O$  is a set of {\em observations} (to be accounted for). We want to identify sets of hypothesis that, in combination with the model, return the observation.

 In the case of the previous example, we have an $\nit{AP}$ with: $H = D^n$, our binary  $M$ is the query program plus the exogenous tuples:\vspace{-2mm}
 {\footnotesize \begin{eqnarray*}
 \nit{yes} &\leftarrow& R(x, y), R(y, z), S(x, y), T(x,y).\\
 \nit{no} &\leftarrow& \nit{not} \ \ \nit{yes}.\\
 T(a,a) &\leftarrow&
 \end{eqnarray*}}

 \vspace{-2mm}\noindent and $O$ is the atom $\nit{yes}$.
 \cite{eiterLeone} provides general algorithms for producing ASPs from $\nit{AP}$s that solve the abductive task; in our case, that return MSSs or mSSs.

In general, the complexity of computational tasks related to abductive problems in ASP, including those with Datalog models (which contain our CQs as recursion-free programs), fall  in the second-level of the polynomial hierarchy in data complexity \cite{eiterTCS97}. This complexity matches the intrinsic complexity of computing minimum SEs \cite{hu}. However, under graph-theoretic conditions on the DB, tractability can be achieved \cite{pichler}.


\vspace{-3mm}
\section{Discussion and Conclusions}\label{sec:disc}

\vspace{-2mm}
\paragraph{\bf 1. On Endogenous and Exogenous Tuples.} To give an example of how these two kinds of tuples may appear in a DB setting, consider a physical or virtual data integration system. Tuples that belong to a data source that is beyond our control or scope of interest could be seen as exogenous. This could be a source that we can treat only as a black box, possibly only via QA. Instead, we may be interested in investigating a different, particular data source, e.g. in terms of its tuples' contributions to QA. Its tuples would be declared as endogenous.

\vspace{-2mm}
\paragraph{\bf 2. \ Necessity Degree vs Sufficiency Degree.}
One could propose the sufficien-cy-degree
as the right metric to measure a tuple contribution to QA. However, the next example, taken from \cite{babakThesis}, suggests that, in some cases, the necessity degree
may give more intuitive results than the sufficiency degree.

\vspace{-2mm}
\begin{example} \label{exa:newcr1}  Consider the instance with relation $E^\prime = \{(t_1,s, a), (t_2, s,b),$ $(t_3,a,c), (t_4,b,c), (t_5,c,t)\}$ corresponding to the graph $\mc{G}^\prime$ below.  Assume $t_1$ and $t_2$, with dashed arrows in $\mc{G}^\prime$, are exogenous. As before, the  query $\mc{Q}$ asks if there is a path from $s$ to $t$.

 \vspace{-4mm}
\begin{multicols}{2}

It holds $E^\prime \models \mc{Q}$, and the positive answer has the following MSSs: $S_1=\{t_2, t_5\}$,
$S_2=\{t_4, t_5\}$, and the following MNSs: $N_1=\{t_5\}$, $N_2=\{t_2, t_4\}$. We obtain: \
$\eta(t_5)=1$, \ $\eta(t_2)=\eta(t_4)= 1/2$, \ but
 $\sigma(t_2)=\sigma(t_4)=\sigma(t_5)= 1/2$.
 
\centerline{\includegraphics[width=1.3in]{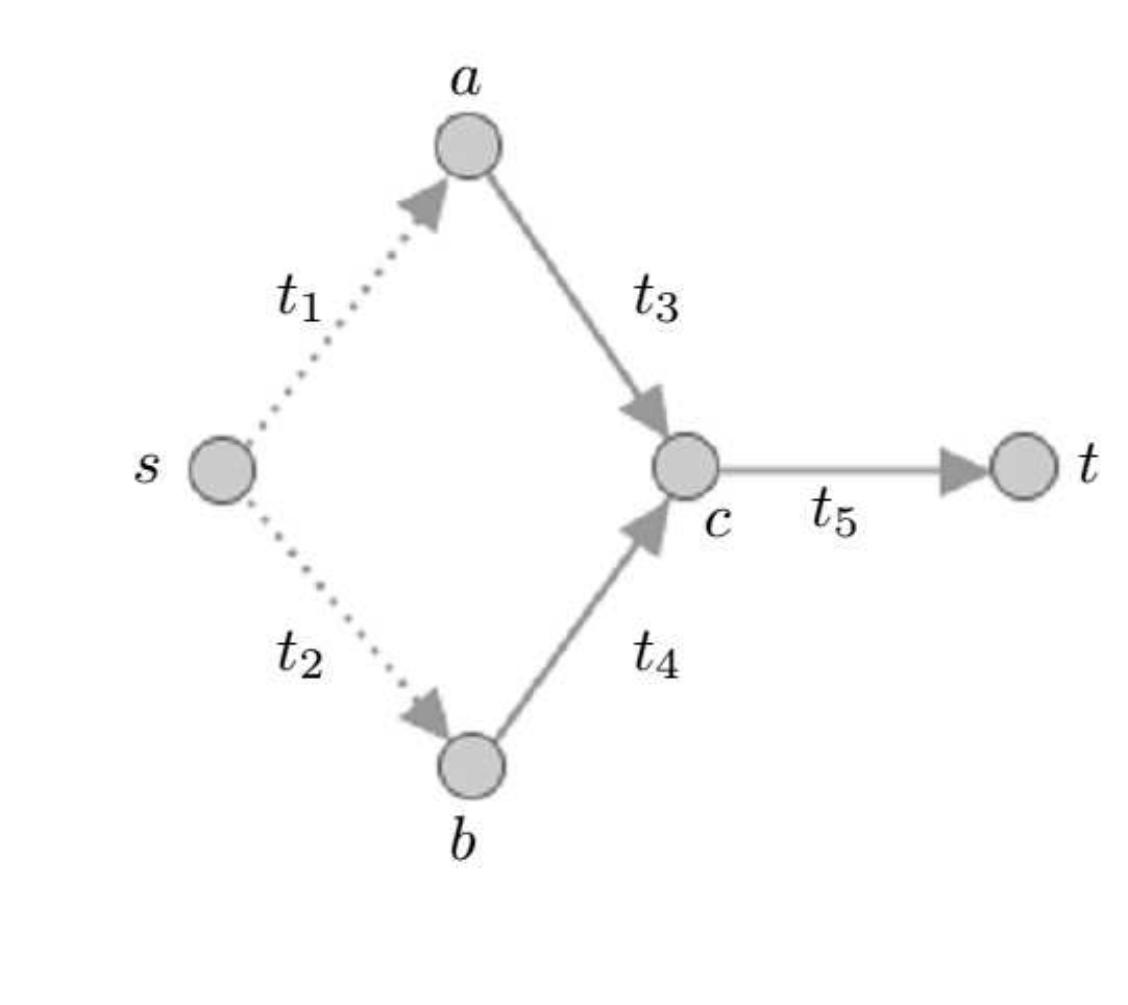}}
 
\end{multicols}

\vspace{-7mm}Intuitively, it makes sense to claim that $t_5$ contributes more to the result, because the answer counterfactually depends on $t_5$ (no need for an accompanying contingency set). 
In this case, in contrast to Example \ref{ex:exIntro3}, ranking causal tuples according to their necessity-degrees is in line with our intuition that they
contribute more to the query result.
\boxtheorem\end{example}

In the light of the examples in this work, the {\em causal-effect score} was introduced as an attribution scores for QA \cite{tapp16}. \ignore{That score has been generalized and further investigated in \cite{clear25,foiks26}.}

In a related but different direction, it would be interesting to investigate under what conditions on the query and the database (e.g. having or not exogenous tuples) the sufficiency and necessity scores are aligned or not. This analysis was carried out for causality-based scores in \cite{foiks26}. Beyond this analysis, it would also be interesting to determine for what kinds of queries and DBs a particular score is more appropriate.  This kind of analysis requires a deeper reflection about explanations for a query result.

\vspace{-2mm}
\paragraph{\bf 3. \ Sufficiency and Circumscribed Lineage.}  The {\em lineage} of a Boolean query $\mc{Q}$  relative to a given instance $D$ is a propositional formula $\mc{L}(\mc{Q},D)$ that becomes true under a truth-valuation determined by $D$ iff  $\mc{Q}$ in $D$ \cite{suciuPDB,tapp16}.

\vspace{-1mm}
\begin{example}  (ex. \ref{ex:suffDB} cont.) \label{ex:lin} The lineage of query $\mc{Q}$ on instance $D^\prime$ is the propositional formula \
$\mc{L}(\mc{Q},D^\prime) := (X_{S(c)} \wedge X_{R(c,b)} \wedge X_{S(b)}) \vee (X_{S(b)} \wedge X_{R(b,b)})$. \
Each $X_t$ is a propositional variable associated to a tuple $t$, with the intended meaning that $X_t$ is true iff $t$ appears in $D^\prime$.  
\boxtheorem\end{example}

\vspace{1mm}
Clearly the elements of $\nit{MSS}(\mc{Q},D)$ correspond to the {\em minimal propositional models} of  $\mc{L}(\mc{Q},D)$; or, in other words, the models of the {\em propositional circumscription} \cite{vlad94,kolaitis04} of $\mc{L}(\mc{Q},D)$ that minimize (with respect to set inclusion) the set of propositional variables that are true. This is what is usually called a {\em parallel circumscription} \cite{vlad94}, i.e. of all variables simultaneously.

 When $D$ is partitioned into $D^\nit{x}\cup D^\nit{n}$, variables $X_t$ associated to exogenous tuples $t$ are not subject to minimization, instead they stay {\em fixed} \cite{vlad94}. In the case of BCQSm, we may assume they are true, and can be eliminated from the query lineage. In Example \ref{ex:lin}, if $R$-tuples are exogenous, the lineage becomes: \  $\mc{L}^\prime(\mc{Q},D^\prime) := (X_{S(c)} \wedge X_{S(b)}) \vee (X_{S(b)})$. \ Notice that for a BCQ the lineage becomes a formula in {\em Monotone Disjunctive Normal Form} (Mon-DNF) in that all literals are positive. Furthermore, for a fixed BCQ $\mc{Q}$ with $d$ atoms, each conjunct of its lineage has at most $d$ variables. \ignore{Recent results in \cite{cooper25} may be applicable under this formulation.}

This lineage-based approach to sufficiency may be suitable for the application of knowledge compilation and model counting techniques, which have turned out to be useful in database research \cite{kumarSuciu}. In this case, the query lineage for a given instance turns out to be a propositional formula in positive DNF with bounded conjunctions.

\ignore{
It is worth mentioning that a circumscriptive specification and computation of database repairs was presented in \cite{bertossiSchwind}; and a circumscription-related specification of database repairs for consistent query answering \cite{CQA} has been proposed \cite{bertossi21}.}



\vspace{-3mm}
 \paragraph{\bf 4. \ Some Additional Research Directions.} Several open problems have been left open and they are a matter of ongoing and future research. 
 We should go beyond conjunctive queries without self-joins, and also into the realm of more general queries, monotone or not. An important direction is to identify conditions on the query that would allow extending Corollary~\ref{cor:sjf+} to the computation of a minimum SS. 

 Another interesting question concerns the impact of the tree-width conditions on the DB, as in \cite{pichler}, on the computation of sufficient explanations and sufficiency degrees. This has been only partially explored in \cite{flairs17} for necessary explanations. Having exogenous tuples may add an interesting dimension to this problem.

 Since the connection between necessary-explanations and repairs has provided much insight on the former, there seems to be ample room for further exploiting the sufficient-explanation and repairs connection. Beyond that, it would be interesting to investigate the repair core in the broader context of {\em model-based diagnosis} \cite{HBKR}.

\vspace{2mm}
\noindent {\bf Acknowledgements:} \ L. Bertossi has been financially supported by the IMFD,
Chile; and NSERC-DG 2023-04650, Canada. N. Pardal has been supported by the Huawei Strategic talent program.

\vspace{-3mm}
\bibliographystyle{plain}
\bibliography{bibtex}

\newpage

\section*{Appendix: Proofs of Results}


\defproof{Proposition~\ref{pro:causesuffset}}{
    We first show the proof for~(\ref{itm:causesuffset}). 
    Assume that $\nit{MSS}=\{S_1,   \ldots,  S_m\}$ contains all MSSs for $\bar{a}$,  and there exists $S \in\nit{MSS}$ with $t \in S$. Consider an arbitrary  $\Gamma \subseteq D^n$ such that, for every $S_i \in \nit{MSS}$ s.t. $S_i \not = S$  it holds  $\Gamma \cap S_i \neq \emptyset$ and  $\Gamma \cap S =\emptyset$. Then, $N=\Gamma \cup \{ t\}$ is a necessary  set for $\bar{a}$.
    It is enough to prove that such a $\Gamma$ always exists. In fact, since all members of $\nit{MSS}$ are subset-minimal,  for each $S_i \in \nit{MSS}$, $S_i \not = S$, it holds $S_i \smallsetminus S \neq \emptyset$. Therefore,  we can choose 
    $\Gamma := \bigcup_1^m \{ \bar{t} \ | \ \bar{t} \in (S_i  \smallsetminus S) \}$. 
    The proof in the other direction is straightforward. 
    
    Now, the statement in~(\ref{itm:sigma-implies-eta}) is a straightforward consequence of~\ref{itm:causesuffset} and the definition of necessary tuple and sufficient explanation.
    
    Finally, we prove that $S \in \nit{MSS}$ iff it is a MHS for $\nit{MNS}$; the proof of the other statement in~(\ref{pro:hitting}) is symmetric.
    %
    Let $S \in \nit{MSS}$. We first show that $S$ is a minimal hitting-set of $\nit{MNS}$.
    Let $N \in \nit{MNS}$. Toward a contradiction, assume that
    $S \cap N = \emptyset$. Then $S \subseteq D^n \smallsetminus N$, and by monotonicity $D^n \smallsetminus N \cup D^x \models \mc{Q}$, contradicting that $N \in \nit{MNS}$. Therefore, $S \cap N \neq \emptyset$ for every $N \in \nit{MNS}$, i.e., $S$ is a hitting-set of $\nit{MNS}$.
    Now let $S' \subsetneq S$. Since $S$ in $\nit{MSS}$, we have $S' \cup D^x \not\models \mc{Q}$.
    Consider the family
    $\mathcal{N}_{S'} := \{ N \subseteq D^n \smallsetminus S' \mid D^n \smallsetminus N \cup D^x \not\models \mc{Q}\}$.
    This family is non-empty (e.g., $D^n \smallsetminus S' \in \mathcal{N}_{S'}$ since $D^n \smallsetminus (D^n \smallsetminus S') = S'$). 
    As $D^n$ is finite, there is a subset-minimal element $N$ of $\mathcal{N}_{S'}$.
    Then $N \in \nit{MNS}$ and, by construction, $N \cap S' = \emptyset$. Hence $S'$ is not a hitting-set of $\nit{MNS}$ and therefore, $S$ is a minimal hitting-set of $\nit{MNS}$.
    
    Conversely, let $S$ be a minimal hitting-set of $\nit{MNS}$. We show that
    $S \in \nit{MSS}$.
    First, suppose that $S \cup D^x \not\models \mc{Q}$, and consider the family
    $\mathcal{N}_{S} := \{ N \subseteq D^n \smallsetminus S \mid D^n \smallsetminus N \cup D^x \not\models \mc{Q} \}$.
    This family is non-empty, since $D^n \smallsetminus S \in \mathcal{N}_{S}$, and thus it contains a a subset-minimal element $N$. Then $N \in \nit{MNS}$ and $N \cap S = \emptyset$, contradicting that $S$ is a hitting-set of $\nit{MNS}$. Therefore, $S \cup D^x \models \mc{Q}$.
    It remains to prove that $S$ is subset-minimal with this property. Let
    $S' \subsetneq S$. If $S' \cup D^x \models \mc{Q}$, then by the already proved implication
    $(\Rightarrow)$, $S'$ would be a hitting-set of $\nit{MNS}$, contradicting the
    minimality of $S$ as a hitting-set. Hence $S' \cup D^x \not\models \mc{Q}$ for every
    proper subset $S' \subsetneq S$, therefore $S \in \nit{MSS}$.}

\defproof{Proposition~\ref{prop:repairs-mhs}}{
If $N$ is in $\mathsf{MHS}(\mc{SH})$, then in particular $N$ intersects every MSS; therefore removing $N$ from $D$ deletes at least one tuple from every witness, so $D\smallsetminus N \not\models\mc{Q}$. Minimality of $N$ as a hitting set matches minimality as a deletion set (no proper subset of $N$ deletes a tuple from every MSS), hence $D\smallsetminus N$ is a repair obtained by a minimal deletion. Conversely, any minimal deletion-set that turns $\mc{Q}$ false must meet every MSS, i.e. it is a hitting set; minimality turns it into a minimal hitting set.}

\defproof{Lemma~\ref{lem:core-exogenous-decomposition}}{
Since $\core$ is the intersection of all repairs, and exogenous tuples cannot be removed, for every repair $R$ of $D$ we have $D^x \subseteq R$. Hence, $\core(D,\mc{K})
= D^x \cup \bigcap_{R \in \nit{Rep}(D,\mc{K})} (R \cap D^n)$.
We prove that $\bigcap_{R \in \nit{Rep}(D,\mc{K})} (R \cap D^n) = \core(D^n,\mc{K})$.

Let $t \in \bigcap_{R \in \nit{Rep}(D,\mc{K})} (R \cap D^n)$.  
Then $t \in D^n$ and $t \in R$ for every $R \in \nit{Rep}(D,\mc{K})$.  
By assumption, for every repair $R$ of $D$, the set $R \cap D^n$ is a repair of $D^n$.  
Hence $t$ belongs to every repair of $D^n$, and therefore $t \in \core(D^n,\mc{K})$.

Let $t \in \core(D^n,\mc{K})$, that is, $t$ belongs to every repair of $D^n$.
Let $R \in \nit{Rep}(D,\mc{K})$. By assumption, $R \cap D^n$ is a repair of $D^n$, and thus $t \in R \cap D^n$ and in particular $t \in R$. Since $R$ was arbitrary, $t \in \bigcap_{R \in \nit{Rep}(D,\mc{K})} (R \cap D^n)$.}

\defproof{Proposition~\ref{prop:eta-core-iff}}{
For the proof of (i): by definition, $\eta(t)=0$ iff $t$ is not contained in any MNS. Identifying MNS with MHS of the hypergraph of minimal witnesses $\mc{SH}$, this means $t\notin\bigcup_{N\in\mathsf{MHS}}N$.
Proposition~\ref{prop:core-formula} (or the identity
$\core(D,\mc{K})=D\smallsetminus\big(\bigcup_{N\in\mathsf{MHS}}N\big)$)
then gives the equivalence with $t\in\core(D,\mc{K})$. 
The proof of (ii) follows as a direct consequence of the latter and Proposition~\ref{pro:causesuffset}~(\ref{itm:sigma-implies-eta}).}

\defproof{Lemma~\ref{claim:1_core_rewriting}}{
    Let $t_1, \ldots, t_k \in P_1^D, \ldots, P_k^D$,  such that \  $T := \{t_1, \ldots, t_k\} \models \mc{Q}$. Hence, for every $D'\subseteq D$ such that $T \subseteq D'$, its holds  $D' \models \mc{Q}$, or equivalently, $D' \not\models \mc{K}$.
    That is, $T$ is a MSS for $\mc{Q}$.

    It is easy to see that one minimal way to repair $D$ is to delete from a fixed $P_i^D$ all the tuples in the relation involved in the satisfaction of $\mc{Q}$, for some combination of tuples in each $P_j^D$, $j\neq i$. This is precisely the set $R_i^D$. Notice that $R_i^D$ is a MNS for $\mc{Q}$.

    It follows that $(D\smallsetminus R_i^D) \not \models \mc{Q}$, equivalently, $(D\smallsetminus R_i^D) \models \mc{K}$. Thus, $D' :=  (D\smallsetminus R_i^D)$ is a minimal repair of $D$ \wrt $\mc{K}$.
    Furthermore, by minimality, for every $i \in \{1, \ldots, k \}$, no repair deletes tuples from $P_i^D$ that are not in $R_i^D$. Also, notice $R_i^D$ is a MSS for $\mc{K}$ and $D'$.
    These two facts together finish the proof.}

\defproof{Proposition~\ref{prop:t_suf_exp}}{
     Assume $t \in D\smallsetminus D^\prime$.  First, since $D'$ is a repair, $D' \models \mc{K}$. \ Now, suppose that  $(D' \smallsetminus \core) \cup \{t\}$ is not a sufficient-set for $\mc{Q}$, that is,  $(D' \smallsetminus \core) \cup \{t\} \not\models \mc{Q}$. Equivalently, $(D' \smallsetminus \core) \cup \{t\} \models \mc{K}$.

     Lemma~\ref{claim:1_core_rewriting} tells us that the tuples in $\core$ are precisely those that are not in conflict with $\mc{K}$, more precisely: \  $t_i \in \core$ if and only if there are no $t_1, \ldots, t_k \in D$ \ with \  $\{t_i \}\cup \{ t_j\}_{j\neq i} \models P_1(\bar{x}) \land \ldots \land  P_k(\bar{x})$.
     Therefore, since $(D' \smallsetminus \core) \cup \{t\} \models \mc{K}$, it follows that $D'\cup \{t\} \models \mc{K}$, which contradicts the minimality of $D'$ as a repair.

    Conversely, assume that $(D'\smallsetminus \core) \cup \set{t}$ is a SS for $D$ and $\mc{Q}$, i.e.  $(D'\smallsetminus \core) \cup \set{t} \models \mc{Q}$. By monotonicity of $\mc{Q}$, $D' \cup \set{t} \models \mc{Q}$. Furthermore, since $D'$ is a repair,  $D' \not\models \mc{Q}$. \ This implies that $t \in D\smallsetminus D'$.}

\defproof{Proposition~\ref{prop:t_suf_exp_alg}}{
    The proof of (a) is clear from Proposition \ref{prop:t_suf_exp}.  \ For (b),  let $t \in D\smallsetminus D'$. Since $t \not\in \core$, then there exists at least one combination of tuples in $P_2, \ldots, P_k$ such that the conjunction of $t$ and said tuples satisfies $\mc{Q}$.

    Let $t_{21}, \ldots, t_{k1}; \ldots; t_{2l}, \ldots, t_{kl}$ be all those possible combinations of tuples. \
    If there is one of these lists such that every tuple lies in $D' \smallsetminus \core$, then we finish the proof, since these tuples satisfy $\mc{Q}$ and thus in particular they must coincide in their common variables.

    If instead for each of these lists there is at least one tuple not in $D'\smallsetminus \core$, then in particular each of these tuples are in $D\smallsetminus D'$ (since there cannot be tuples in conflict with other tuples in $\core$). Moreover, we can put $t$ in $D'$ and obtain $D''$ such that $D \supseteq D'' \supsetneq D'$.
    Furthermore, $D''$ is a repair of $D$ since every list of tuples in conflict with $t$ is listed, and for each list there is at least one tuple different from $t$ that was already deleted in $D'$, thus in $D''$, hence contradicting the minimality of $D'$.

    This proves the existence of a MSS in $(D'\smallsetminus \core) \cup \{t\}$. It follows that the procedure defined in the statement of the proposition determines how to find said MSS.}

\end{document}